\documentclass[12pt]{article}
\usepackage{mathbbold}
\usepackage{amsfonts}
\usepackage{amsmath}
\usepackage{amssymb}
\usepackage{mathrsfs}
\usepackage{amstext}
\usepackage{amsfonts}
\usepackage{chemarr,stmaryrd}
\usepackage{graphicx,epsfig}
\usepackage{indentfirst}
\newtheorem{theorem}{Theorem}[section]

\makeatletter
\def\ExtendSymbol#1#2#3#4#5{\ext@arrow 0099{\arrowfill@#1#2#3}{#4}{#5}}
\def\RightExtendSymbol#1#2#3#4#5{\ext@arrow 0359{\arrowfill@#1#2#3}{#4}{#5}}
\def\LeftExtendSymbol#1#2#3#4#5{\ext@arrow 6095{\arrowfill@#1#2#3}{#4}{#5}}
\makeatother

\makeatletter
\renewcommand*\env@matrix[1][*\c@MaxMatrixCols c]{%
  \hskip -\arraycolsep
  \let\@ifnextchar\new@ifnextchar
  \array{#1}}
\makeatother

\begin{document}
\baselineskip 20pt

\title{Classification of the Entangled states $L\times N\times N$}
\author{Jun-Li Li$^1$, Shi-Yuan Li$^2$, and
Cong-Feng Qiao$^{1,3,}$\footnote{Corresponding author: qiaocf@gucas.ac.cn}\\[0.5cm]
$^{1}$Department of Physics, Graduate University of Chinese Academy of Sciences \\
YuQuan Road 19A, Beijing 100049, China\\[0.2cm]
$^{2}$School of Physics, Shandong University, 250100, China\\[0.2cm]
$^{3}$Theoretical Physics Center for Science Facilities (TPCSF),
CAS\\ YuQuan Road 19B, Beijing 100049, China}

\date{}
\maketitle

\begin{abstract}
We presented a general classification scheme for the tripartite
$L\times N\times N$ entangled system under stochastic local
operation and classical communication. The whole classification
procedure consists of two correlated parts: the simultaneous
similarity transformation of a commuting matrix pair into a
canonical form and the study of internal symmetry of parameters in
the canonical form. Based on this scheme, a concrete example of
entanglement classification for a $3\times N\times N$ system is given. \\


\end{abstract}

\section{Introduction}

Entanglement represents the essence of quantum theory compared to
the classical theories. With the development of quantum information
science, entanglement is now expected to be generated from a wide
range of systems and is generally regarded as the key physical
resource in implementing various quantum information tasks
\cite{qcqi}. The classification of entanglement is generally a
mathematically difficult problem. It has been pointed out that only
in a kind of tripartite system, which has a qubit as one of the
subsystems ($2\times M\times N$), one may find finite number of
entanglement classes \cite{Three-qubit}. The complete entanglement
classification of a $2\times M\times N$ pure system under stochastic
local operation and classical communication (SLOCC) has been done in
Refs. \cite{2nn,2mn}, where canonical forms are constructed via
Jordan decomposition of the matrices. There the eigenvalues of the
Jordan forms serve as the continuous parameters in the entanglement
classes \cite{2nn}. Despite the great progress made in $2\times
M\times N$ systems, in many classification schemes the
classification of $3\times N\times N$ turns out to be a ``wild''
problem \cite{Belitskii-Sergeichuk}. Attempts toward the
classification of a multipartite system in literature mainly
concentrate on its tensor ranks or local ranks of the tensor form of
the quantum state, and there is still no systematic method for
constructing the canonical form of $L\times N\times N$ under SLOCC
in literature to our knowledge. Recently, a method featured by the
decomposition of the entanglement classification under local unitary
transformations into different parts has been proposed
\cite{LU-HOSVD}. In this work, by decomposing entanglement
classification into two complementary ingredients, we propose an
entanglement classification scheme for a tripartite $L\times N\times
N$ system under SLOCC.

The structure of the paper is as follows. In Sec. 2, by representing
the $L\times N\times N$ quantum state in the form of a three-way
tensor, its entanglement classification under SLOCC is transformed
into two correlated problems: the simultaneous similarity
transformation of a commutating matrix pair into a canonical form
and the representation of the symmetry of parameters in the
canonical form. In Sec. 3, a concrete example of entanglement
classification for a $3\times N\times N$ system is given in which
the explicit form of the symmetry property of the state is shown. A
brief summary and concluding remarks are given in Sec. 4.

\section{The classification scheme}

Considering a general tripartite pure quantum state with dimensions
$L\times N\times N$, the wave function can be written as
\begin{eqnarray}
\psi = \sum_{i,j,k=1}^{L,N,N} \gamma_{ijk}|i\rangle |j\rangle
|k\rangle \; ,
\end{eqnarray}
where $|i\rangle, |j\rangle ,|k\rangle$ are bases of the three parts
respectively. The complex coefficients $\gamma_{ijk}$ can be treated
as a three-order tensor which then turns into the form of a tuple of
$L$ complex $N\times N$ matrices, i.e.,
\begin{eqnarray}
\psi = (\Gamma_{1},\Gamma_{2},\cdots ,\Gamma_{l}) \; ,
\label{state-form-matrices}
\end{eqnarray}
where $\Gamma_{i}, i\in \{1,2,\cdots ,L\}$ are $N\times N$ complex
matrices and have $[\gamma_i]_{jk}$ as their elements. A state
$\psi' = (\Gamma'_{1}, \Gamma'_{2},\cdots , \Gamma'_{L})$ is said to
be SLOCC equivalent \cite{Three-qubit} to $\psi$ in the case of
\begin{eqnarray}
\sum_{j=1}^{L}T_{ij}P\Gamma_{j}Q = \Gamma'_{i}\; , \; i= \{1,
2,\cdots , L\} \; . \label{TPQ-def}
\end{eqnarray}
Here, $T$ (where its elements are $T_{ij}$) is an $L\times L$
invertible matrix, i.e., invertible local operator (ILO) acting on
the first partite, and $P$ and $Q$ are $N\times N$ ILOs acting on
the second and third partite, respectively.

For a tuple of matrices $(\Gamma_1, \Gamma_2, \cdots,\Gamma_L)$,
there always exists a tuple of numbers $(t_1, t_2, \cdots, t_L)$
$t_i\in \mathbb{C}$, such that the linear combination of
$\Gamma_{i}$s,
\begin{eqnarray}
\Gamma = \sum_{j=1}^{L}t_{j}\Gamma_j  \label{maxRank}
\end{eqnarray}
gives the maximum rank, the $r(\Gamma)$. (We denote the rank of a
matrix as $r(\cdot)$ hereafter.) For every quantum state in the form
of Eq.(\ref{state-form-matrices}), we can always transform them into
the form where the $\Gamma_1$ has the maximum rank by performing the
transformation $T$, of which the first row is chosen to be the tuple
of $(t_1, t_2, \cdots, t_L)$, i.e., $T_{1j}=t_j$. Therefore, in the
following we can then restrict our discussion of the quantum state
classification to that of the tuple $(\Gamma_1,
\Gamma_2,\cdots,\Gamma_L)$, where $\Gamma_1$ has possessed the
maximum rank.

The strategy we adopt in the entanglement classification in this
work is quite transparent, which can be roughly outlined as the
manipulation of three commutative ILOs $T$, $P$, $Q$ in different
order and combination. We first transform the quantum state in a
tuple of matrices according to specific rank property for each
matrix by a subset of ILO $T$, then obtain a canonical form by using
ILOs $P$, $Q$, and finally achieve the entanglement classification
by reconsidering the rest a subset of ILO $T$ as a degeneracy
condition between the canonical forms. Following, we explain in
detail the classification procedures, in which two situations of
matrix $\Gamma_1$ with full and nonfull ranks are discussed
separately.

\subsection{The case of $\Gamma_1$ with full rank} \label{sec-fullR}

In this case, obviously every quantum state in form
(\ref{state-form-matrices}) can be transformed into the following
form by ILOs $P,Q$ defined in Eq.(\ref{TPQ-def})
\begin{eqnarray}
\psi= (E, \Gamma_2,\cdots, \Gamma_L) \; , \label{FullRank}
\end{eqnarray}
where $E$ is a unit matrix. Then a subsequent ILO transformation
\begin{eqnarray}
T = \begin{pmatrix} 1 & 0 & \cdots & 0 \\ -t_{21} & 1 & \cdots & 0
\\ \vdots & \vdots & \ddots & \vdots \\ -t_{L1} & 0 & \cdots & 1
\end{pmatrix} \; ,
\end{eqnarray}
where $t_{i1}$ is one of the eigenvalues of $\Gamma_{i}$, can always
be applied to $\psi$ and makes $r(\Gamma_i)<N$ after the
transformation. By definition, two such quantum states $\psi, \psi'$
are equivalent under the transformations of ILOs $P$, $Q$ if and
only if
\begin{eqnarray}
(PEQ,P\Gamma_2Q,\cdots, P\Gamma_LQ) = (E,\Gamma'_2, \cdots,
\Gamma'_L) \; , \label{PQ-transform}
\end{eqnarray}
which tells $Q=P^{-1}$. Equation (\ref{PQ-transform}) now turns into
a simultaneous similarity relation of matrices $(\Gamma_2,\cdots,
\Gamma_L)$,
\begin{eqnarray}
(PEP^{-1},P\Gamma_2P^{-1}, \cdots, P\Gamma_LP^{-1}) = (E, \Gamma'_2,
\cdots, \Gamma'_L) \; . \label{PQ-similarity}
\end{eqnarray}
According to \cite{Gelfand-Ponomarev, Pairs-annihilating}, all these
kinds of problems can be transformed into the simultaneous
similarity of a pair of commuting matrices, i.e.,
Eq.(\ref{PQ-similarity}) is equivalent to
\begin{eqnarray}
(E,PA_2P^{-1}, PA_3P^{-1}) = (E,A'_2, A'_3) \; ,
\label{PP-Commut-pair}
\end{eqnarray}
where $A_{2,3}$ are composed by $\Gamma_{i}$, and $[A_2,A_3]=0$.
This makes the entanglement classification of a $3\times N\times N$
system of great importance. The canonical form of
Eq.(\ref{PP-Commut-pair}) under the similarity transformation can be
written as $\{ E, J , A \}$, where $J$ is the Jordan form of $A_2$,
$A$ is the canonical form of $A_3$, and $[A,J]=0$. $J$ can be
represented as $J = \bigoplus J_{n_i}(\lambda_i)$, where
$J_{n_i}(\lambda_i)$ are $n_i\times n_i$ matrices which take the
following form:
\begin{eqnarray}
\begin{pmatrix}
\lambda_i & 1 & 0 &\cdots & 0 \\
0 & \lambda_i & 1 & \cdots & 0 \\
0 & 0 & \lambda_i & \cdots & 0 \\
\vdots & \vdots & \vdots & \ddots & 1 \\ 0 & 0 & 0 & \cdots &
\lambda_i
\end{pmatrix} \; .
\end{eqnarray}
Due to $[A,J]=0$, the matrix $A$ generally has the form of $A =
\bigoplus A_{i}$ accordingly, where $A_{i}$ is a $n_i\times n_i$
triangular Toeplitz matrix (see Appendix \ref{structure-matrix-A}):
\begin{eqnarray}
A_{i} = \begin{pmatrix} a_{i0} & a_{i1} & a_{i2} & a_{i3} & \cdots &
a_{in_i-1} \\ 0 & a_{i0} & a_{i1} & a_{i2} & \cdots & a_{in_i-2} \\
0 & 0 & a_{i0} & a_{i1} & \cdots & a_{in_i-3}
\\ 0 & 0 & 0 & a_{i0} & \cdots & a_{in_i-4} \\ \vdots & \vdots & \vdots & &
\ddots & \vdots \\ 0 & 0 & 0& 0 & \cdots & a_{i0}
\end{pmatrix} \; . \label{A_i-form}
\end{eqnarray}
Since the Jordan block can be defined as a polynomial function of
$J_{n_i}(0)$,
\begin{eqnarray}
J_{n_i}(\lambda_i) = \lambda_i + J_{n_i}(0)\; , \label{Jordan-fun}
\end{eqnarray}
the matrix $A_i$ can also be similarly defined as
\begin{eqnarray}
A_i = \sum_{n=0}^{n_i-1} a_{in}J^n_{n_i}(0) \; . \label{origin-Ai}
\end{eqnarray}
By defining $f_{[A_i]}(x) = \sum_{n=0}^{n_i-1} a_{in}x^n$, we have $
A_i = f_{[A_i]}(J_{n_i}(0))$.

However, there still exist other transformation $T$s that can relate
$\psi = (E, \Gamma_2,\cdots, \Gamma_L)$ to another quantum state
$\psi' = (E, \Gamma_2',\cdots, \Gamma_L')$. The effect of the
transformation $T$ is introducing a degeneracy on the canonical form
of $( E, J , A )$. That is, the states $( E, J , A )$ and $( E, J' ,
A' )$ with parametric difference may belong to the same entanglement
class. Define a set $\vec{\mathbf{a}} = \{\lambda_i, a_{in}|0\leq n
\leq n_i-1\}$, which contains all the parameters in $( E, J , A )$.
The degeneracy can then be expressed as a symmetry transformation
$\mathcal{P}$ on the parameters of the canonical form, i.e.,
\begin{eqnarray}
\vec{\mathbf{a}}\,' = \mathcal{P} \,\vec{\mathbf{a}} \; .
\end{eqnarray}
Thus, the canonical form is of the entanglement classification of
the quantum state up to the symmetry $\mathcal{P}$.

\subsection{The case of $\Gamma_1$ with non-full rank}

In this case, all of the quantum states can be transformed into the
following form by ILOs $P,Q$
\begin{eqnarray}
\psi= (\Lambda, \Gamma_2,\cdots, \Gamma_L ) \; . \label{LessRank}
\end{eqnarray}
Here, $\Lambda$ has the following form:
\begin{eqnarray}
\Lambda =
\begin{pmatrix}
E_{(N-i)\times (N-i)} & \mathbf{0} \\
\mathbf{0} & \mathbf{0}_{i\times i}
\end{pmatrix} =
\begin{pmatrix}[ccc|ccc] 1 & \cdots & 0 & 0 & \cdots & 0 \\
\vdots & \ddots & \vdots  & \vdots & \ddots & \vdots \\ 0 & \cdots &
1 & 0 &
\cdots & 0 \\ \hline 0 & \cdots & 0 & 0 & \cdots & 0 \\
\vdots & \ddots & \vdots  & \vdots & \ddots & \vdots \\ 0 & \cdots &
0 & 0 & \cdots & 0
\end{pmatrix} \;
\end{eqnarray}
with $i>0~$.

Then we apply another type of ILOs, $P_{\Lambda}, Q_{\Lambda}$, such
that they keep $\Lambda$ invariant and transform all the $\Gamma_j$,
$j\in \{2,\cdots, L\}$ to a form of direct sum of two matrices. The
philosophy here is the same as that of the less than full rank case
in \cite{2nn}, that is,
\begin{eqnarray}
\Gamma_j = \begin{pmatrix}[ccc|ccc] \gamma_{j11} & \cdots &
\gamma_{j1n} & 0 & \cdots & 0
\\ \vdots & \ddots & \vdots  & \vdots & \ddots & \vdots \\ \gamma_{jn1} &
\cdots & \gamma_{jnn} & 0 &
\cdots & 0 \\ \hline 0 & \cdots & 0 & \beta_{j11} & \cdots & \beta_{j1m} \\
\vdots & \ddots & \vdots  & \vdots & \ddots & \vdots \\ 0 & \cdots &
0 & \beta_{jm1} & \cdots & \beta_{jmm}
\end{pmatrix} \; . \label{partition-Gamma_j}
\end{eqnarray}
Here, $N=n+m$, and the matrices $\beta_j$ cannot be simultaneously 0
for $j$ and cannot be further decomposed into the direct sum of
submatrices via $P_{\Lambda},Q_{\Lambda}$. From this and from the
condition that the first matrix $\Lambda$ has the maximum rank in
comparison to $\Gamma_{\!j}$, we can also deduce $m>i$. With the
partition (\ref{partition-Gamma_j}), $\Lambda$ can be partitioned
accordingly as
\begin{eqnarray}
\Lambda =\begin{pmatrix} E_{n\times n} & 0 \\ 0 & \Lambda'_{m\times
m} \end{pmatrix} =\begin{pmatrix}[ccc|ccc] 1 & \cdots & 0 & 0 &
\cdots & 0
\\ \vdots & \ddots & \vdots  & \vdots & \ddots & \vdots \\ 0 &
\cdots & 1 & 0 &
\cdots & 0 \\ \hline 0 & \cdots & 0 & 1 & \cdots & 0 \\
\vdots & \ddots & \vdots  & \vdots & \ddots & \vdots \\ 0 & \cdots &
0 & 0 & \cdots & 0
\end{pmatrix} \; , \label{partition-Lambda}
\end{eqnarray}
where
\begin{eqnarray}
\Lambda' = \begin{pmatrix}
E_{(m-i)\times (m-i)} & \mathbf{0} \\
\mathbf{0} & \mathbf{0}_{i\times i}
\end{pmatrix}.
\end{eqnarray}

Now from the direct sum forms of Eqs.(\ref{partition-Gamma_j},
\ref{partition-Lambda}), the task of classification under
$P_{\Lambda},Q_{\Lambda}$ can be divided into two problems: the
simultaneous similarity transformation of matrices $\gamma_j$ and
the construction of matrices $\beta_j$ in canonical form. The first
one can be solved with the help of results in Sec. \ref{sec-fullR}.
For the second problem the necessary and sufficient condition for
the canonical form of $\beta_j$ goes as follows:
\begin{eqnarray}
r_{\mathrm{max}}(\Lambda' + \sum_{j=2}^{L}\alpha_j\beta_j)= m-i \, ,
\, \alpha_j \in \mathbb{C} \; .
\end{eqnarray}

Similar to the full rank situation, the transformation $T$ here will
also induce a symmetry $\mathcal{P}$ to the canonical form, which
can be schematically expressed as
\begin{eqnarray} \vec{~\mathbf{b}}'
= \mathcal{P} \vec{~\mathbf{b}} \; ,
\end{eqnarray}
where $\vec{~\mathbf{b}} = (\lambda_i, a_{in}, \beta_{jkl})$.

In all one can conclude that for each matrix triple $(E,J,A)$, the
following theorem holds:
\begin{theorem}
The classification of $L\times N\times N$ can be represented in a
triple-matrix form $(E,J,A)$ up to a parametric symmetry
$\mathcal{P}$.
\end{theorem}

Now we note that the entanglement classification of quantum state $L
\times N\times N$ under SLOCC can be decomposed into two problems:
the simultaneous similarity transformation of a commutating matrix
pair and the symmetry analysis of the corresponding canonical form.
Therefore the difficulty in entanglement classification then turns
to that in the symmetry property study, which requires further
analysis for different entanglement structures. Analysis of the
symmetry properties represents the essence of the entanglement class
in our classification scheme. In the following we show the explicit
form of symmetry in the classification of the $3\times N\times N$
quantum system, from which the symmetry nature of state $L\times
N\times N$ can be figured out.

\section{Entanglement classification of  $3\times N\times N$
system}\label{sec-3NN}

In this section we give an explicit example of a $3\times N\times N$
system: $\psi = (A_1, A_2, A_3)$, where $r(A_1) = N$ (full-rank
case), and when transformed into $(E, A'_2, A'_3)$ via $P,Q$ it has
$[A'_2, A'_3] = 0$. From the discussion in Sec. 2 it is clear that
the canonical form of this state is $(E,J, A)$, where $J$ is the
Jordan form of $A'_2$ and $[J,A] = 0$. We can always achieve
$r(E)>r(J)>r(A)$ by applying a certain transformation $T$. In order
to realize the unique entanglement classification, we need to find
out the specific form of symmetry $\mathcal{P}$.

An invertible matrix $T$ can be decomposed as $T=\mathbb{P}LDU$,
where $L$ $(U)$ is the lower (upper) triangular with all diagonal
entries equal to 1, $D$ is a nonsingular diagonal matrix, and
$\mathbb{P}$ is a permutation matrix \cite{matrix-analysis}. Among
these four matrices $\mathbb{P},L,D,U$: (1) $\mathbb{P}$ and $L$ are
now contradicting with $r(E) >r(J)> r(A)$ in most cases, except for
very limited (less than the number of distinct values of
$\lambda_i$) and isolated values which can be treated individually;
and (2) $D$ corresponds to a trivial rescale of the whole matrix
which can be treated by the renormalization of quantum state. Now
the remaining effective $T$ transformation is
\begin{eqnarray}
U = \begin{pmatrix}
1 & z_{1} & z_{2} \\
0 & 1 & z_{3} \\
0 & 0 & 1
\end{pmatrix} =
\begin{pmatrix}
1 & z_{1} & 0 \\
0 & 1 & 0 \\
0 & 0 & 1
\end{pmatrix}
\begin{pmatrix}
1 & 0 & z_{2} \\
0 & 1 & 0 \\
0 & 0 & 1 \end{pmatrix}
\begin{pmatrix}
1 & 0 & 0 \\
0 & 1 & z_{3} \\
0 & 0 & 1
\end{pmatrix} \; . \label{T=PLDU}
\end{eqnarray}
The admissible $T$ matrix which can be applied on the canonical form
$(E,J, A)$ is now decomposed into a sequence of three operations,
like
\begin{eqnarray}
\begin{pmatrix}E \\ J\\ A \end{pmatrix}
\xrightarrow{T^{(z_1)}_{[EJ]}}
\begin{pmatrix}E + z_1 J\\ J\\ A \end{pmatrix} ,
\begin{pmatrix}E \\ J\\ A \end{pmatrix}
\xrightarrow{T^{(z_2)}_{[EA]}}
\begin{pmatrix}E + z_2 A\\ J \\ A \end{pmatrix} ,
\begin{pmatrix}E \\ J \\ A \end{pmatrix}
\xrightarrow{T^{(z_3)}_{[JA]}}
\begin{pmatrix}E \\ J + z_3A \\ A \end{pmatrix} \; ,
 \label{superposition-EJA}
\end{eqnarray}
where $T^{(z)}_{[XY]}$ is an elementary operation that multiplies
$Y$ with $z$ and adds it to the $X$ column. Because all the $E,J,A$
are direct sums of block-diagonalized submatrices and have the same
partitions, we can therefore treat Eq.(\ref{superposition-EJA}) with
their diagonalized blocks.

\subsection{The first superposition $T^{(z_1)}_{[EJ]}$}\label{TzEJ}

After the transformation $T^{(z_1)}_{[EJ]}$, the initial canonical
form turns to
\begin{eqnarray}
\begin{pmatrix}E + z_1 J\\ J\\ A \end{pmatrix} \; ,
\end{eqnarray}
which is no longer a canonical form. To obtain the canonical form
corresponding to it, operators $P$ and $Q$ should satisfy
\begin{eqnarray}
P( E + z_1 J_{n_i}(\lambda_i))Q = E\; ,
\end{eqnarray}
for each block. This equation can be reexpressed as
\begin{eqnarray}
P g^{(z_1)}_{\scriptscriptstyle[EJ]}(J_{n_i}(0))Q = E \; ,
\label{gea-pq}
\end{eqnarray}
where $g^{(z_1)}_{\scriptscriptstyle[EJ]}(x) = 1 + z_1\lambda_i +
z_1 x$. The reciprocal of polynomial function
$g^{(z_1)}_{\scriptscriptstyle[EJ]}(x)$ can be readily obtained,
that is (note $J^{n}_{n_i}(0)=0$ if $n\geq n_i$),
\begin{eqnarray}
\zeta^{(z_1)}_{\scriptscriptstyle[EJ]}(J_{n_i}(0)) =
\left[g^{(z_1)}_{\scriptscriptstyle[EJ]}(J_{n_i}(0))\right]^{-1} =
\frac{1}{1 +z_1 \lambda_i} \sum_{n=0}^{n_i} \left(\frac{-z_1 }{1
+z_1 \lambda_i}\right)^nJ^n_{n_i}(0) \; .
\end{eqnarray}
Since Eq.(\ref{gea-pq}) leads to $ P = Q^{-1} \cdot
\zeta^{(z_1)}_{\scriptscriptstyle[EJ]}(J_{n_i}(0))$, we can further
get
\begin{eqnarray}
PJ_{n_i}(\lambda_i)Q  & = & Q^{-1} \cdot
\zeta^{(z_1)}_{\scriptscriptstyle[EJ]}(J_{n_i}(0)) \cdot
J_{n_i}(\lambda_i) \cdot Q \nonumber \\ & = & Q^{-1} \cdot \left[
\frac{\lambda_i}{1 + z_1\lambda_i} + \sum_{n=1}^{n_i} \frac{1}{(1
+z_1\lambda_i)^2} \left(\frac{-z_1}{1 +
z_1\lambda_i}\right)^{n-1}J^n_{n_i}(0) \right] \cdot Q \nonumber \\
& = & Q^{-1} \cdot \, f^{(z_1)}_{[(EJ)J]}(J_{n_i}(0)) \cdot \, Q\;\;
. \label{1PQ-J}
\end{eqnarray}
Here $f^{(z_1)}_{[(EJ)J]}(J_{n_i}(0))$ is defined to be
$\zeta^{(z_1)}_{\scriptscriptstyle[EJ]}(J_{n_i}(0)) \cdot
J_{n_i}(\lambda_i)$, which represents the influence of the
superposition of $E$ and $z_1 J$ on $J$ after the operation $P$,
$Q$. Similarly, we can obtain this operation influence on $A_i$,
i.e.,
\begin{eqnarray}
PA_iQ & = & P \cdot f_{[A_i]}(J_{n_i}(0)) \cdot Q \nonumber \\
& = & Q^{-1}\cdot\,
\zeta^{(z_1)}_{\scriptscriptstyle[EJ]}(J_{n_i}(0))\cdot
f_{[A_i]}(J_{n_i}(0)) \cdot\,Q \nonumber
\\ & = & Q^{-1}\cdot \, \left[ \sum_{n=0}^{n_i}
\frac{1}{1 +z_1\lambda_i} \sum_{m=0}^{n} \left(\frac{-z_1}{1
+ z_1\lambda_i}\right)^{m}a_{in-m}
J_{n_i}^{n}(0) \right] \cdot\, Q \nonumber \\
& = & Q^{-1} \cdot \, f^{(z_1)}_{[(EJ)A]}(J_{n_i}(0)) \cdot \,Q \;\;
.
\end{eqnarray}
In order to see what the canonical form becomes now, we perform a
similarity transformation $S$ on the second matrix,
Eq.(\ref{1PQ-J}), and transform it to the Jordan form
\begin{eqnarray}
S^{-1}Q^{-1} \cdot\, f^{(z_1)}_{[(EJ)J]} (J_{n_i}(0)) \cdot \,QS =
J_{n_i}\left(\frac{\lambda_i}{1 +z_1\lambda_i}\right) \; ,
\label{SQJQS}
\end{eqnarray}
which can be verified with the help of (\ref{1PQ-J}). From
(\ref{SQJQS}) we then have
\begin{eqnarray}
S^{-1}Q^{-1}\cdot \, J_{n_i}(0)\cdot\, QS = f_{[(EJ)J]}^{{(z_1)}-1}
\left[ J_{n_i} \left( \frac{\lambda_i}{1 + z_1\lambda_i} \right)
\right] \; .
\end{eqnarray}
Here $f_{[(EJ)J]}^{{(z_1)}-1}$ is the inverse function of
$f_{[(EJ)J]}^{{(z_1)}}$. This similarity transformation will also
make
\begin{eqnarray}
S^{-1}Q^{-1} f^{(z_1)}_{[(EJ)A]} (J_{n_i}(0)) QS & = &
f^{(z_1)}_{[(EJ)A]} \left( S^{-1}Q^{-1} J_{n_i}(0) QS \right)
\nonumber \\ & = & f^{(z_1)}_{[(EJ)A]} \circ f_{[(EJ)J]}^{(z_1)-1}
\circ J_{n_i} \left( \frac{\lambda_i}{1 + z_1\lambda_i} \right)\; ,
\end{eqnarray}
where the Jordan canonical form is treated as a polynomial function
[see Eq.(\ref{Jordan-fun})] and $f\circ g(x) \equiv f(g(x))$.

We note that when the $T$ transformation of $ E + z_1 J$ applies on
\begin{eqnarray}
(E,J_{n_i}(\lambda_i),A_{i}) \; , \label{EJAbefore121}
\end{eqnarray}
where $A_i = f_{[A_i]}(J_{n_i}(0))$, it will make the canonical form
transform into
\begin{eqnarray}
(E,J_{n_i}\left( \frac{\lambda_i}{1 + z_1\lambda_i} \right),A'_{i})
\; , \label{EJAafter122}
\end{eqnarray}
where $A'_i = f^{(z_1)}_{[(EJ)A]} \circ f_{[(EJ)J]}^{(z_1)-1} \circ
J_{n_i} \left( \displaystyle \frac{\lambda_i}{1 + z_1\lambda_i}
\right)$. Though (\ref{EJAbefore121}) and (\ref{EJAafter122}) are
similar in block partitions with different parameters, they are
equivalent entanglement classes. That means the entanglement classes
parameterized by $\vec{\mathbf{a}} = (\lambda_i, a_{in})$ are not
inequivalent with different parametric values, while parameters
undergo a transformation $\mathcal{P}_{z_1}$ characterized by $z_1$,
i.e.,
\begin{eqnarray}
\left.\begin{array}{l} \lambda_i \\  \\ a_{in} \end{array}\right\}
\xrightarrow{\displaystyle \mathcal{P}_{z_1}} \left\{
\begin{array}{l} \displaystyle \frac{\lambda_i}{1 + z_1\lambda_i}
\\ \\ a'_{in}
\end{array} \right. \; ,\label{first-superposition}
\end{eqnarray}
where $a_{in}'$ is the entries of $A_{i}'$ defined in
Eq.(\ref{EJAafter122}), or simply,
\begin{eqnarray}
\begin{pmatrix} E \\ J' \\ A' \end{pmatrix} =
\mathcal{P}_{z_1} \begin{pmatrix} E\\ J\\ A
\end{pmatrix} \; .
\end{eqnarray}

\subsection{The second superposition $T^{(z_2)}_{[EA]}$}
\label{sec-TEA}

Similar to the above section, the transformation requirement
\begin{eqnarray}
P( E + z_2 A_{i})Q = E
\end{eqnarray}
leads to
\begin{eqnarray}
P \cdot \, g^{(z_2)}_{\scriptscriptstyle[EA]}\left[J_{n_i}(0)\right]
\cdot \, Q = E \; ,
\end{eqnarray}
where the polynomial $ g^{(z_2)}_{\scriptscriptstyle[EA]}(x) =
\sum_{n=0}a_{n}x^n $, with $a_0 = 1 + z_2 a_{i0}$, $a_{j}= z_2
a_{ij}$ when $j>0$. Define
$\zeta^{(z_2)}_{\scriptscriptstyle[EA]}(J_{n_i}(0)) =
[g^{(z_2)}_{\scriptscriptstyle[EA]}(J_{n_i}(0))]^{-1}$ such that $P
= Q^{-1}\cdot \, \zeta^{(z_2)}_{\scriptscriptstyle[EA]}
(J_{n_i}(0))$. What we need to know now are the values of
$PJ_{n_i}(\lambda_i)Q$ and $P A_i Q$.

For $P \, J_{n_i}(\lambda_i) \, Q$, we have
\begin{eqnarray}
PJ_{n_i}(\lambda_i)Q  & = & Q^{-1}\cdot \,
\zeta^{(z_2)}_{\scriptscriptstyle[EA]}(J_{n_i}(0)) \cdot
J_{n_i}(\lambda_i) \cdot \, Q \nonumber  \\ & = & Q^{-1}\cdot \,
f^{(z_2)}_{[(EA)J]}(J_{n_i}(0)) \cdot \, Q\; .
\end{eqnarray}
Here $f^{(z_2)}_{[EAJ]}(J_{n_i}(0))\equiv
\zeta^{(z_2)}_{\scriptscriptstyle[EJ]}(J_{n_i}(0)) \cdot
J_{n_i}(\lambda_i)$. For $P A_i Q$, we have
\begin{eqnarray}
PA_iQ & = & P \cdot \, f_{[A_i]}(J_{n_i}(0)) \cdot \,Q \nonumber \\
& = & Q^{-1}\cdot\,
\zeta^{(z_2)}_{\scriptscriptstyle[EA]}(J_{n_i}(0))\cdot
f_{[A_i]}(J_{n_i}(0)) \cdot \, Q \nonumber
\\  & = &
Q^{-1}\cdot \,f^{(z_2)}_{[(EA)A]}(J_{n_i}(0)) \cdot \, Q \; .
\end{eqnarray}
There always exists a similarity transformation such that
\begin{eqnarray}
& & S^{-1}Q^{-1} f^{(z_2)}_{[(EA)J]} (J_{n_i}(0)) QS = J_{n_i}\left(
\frac{\lambda_i}{1 + z_2 a_{i0}} \right)
\end{eqnarray}
and hence
\begin{eqnarray}
S^{-1}Q^{-1}J_{n_i}(0)QS = f^{(z_2)-1}_{[(EA)J]} \left[ J_{n_i}
\left( \frac{\lambda_i}{1 + z_2 a_{i0}} \right) \right] \; .
\end{eqnarray}
Then we have
\begin{eqnarray}
& & S^{-1}Q^{-1} f^{(z_2)}_{[(EA)A]} (J_{n_i}(0)) QS \nonumber \\  &
= &
f^{(z_2)}_{[(EA)A]}\left( S^{-1}Q^{-1} J_{n_i}(0) QS \right) \nonumber \\
& = & f^{(z_2)}_{[(EA)A]}\circ f^{(z_2)-1}_{[(EA)J]} \circ J_{n_i}
\left( \frac{\lambda_i}{ 1+ z_2 a_{i0}} \right) \; .
\end{eqnarray}
We note that when the $T$ transformation of $ E + z_2 A_i$ applies
on
\begin{eqnarray}
(E,J_{n_i}(\lambda_i),A_{i}) \; ,
\end{eqnarray}
where $A_i = f_{[A_i]}(J_{n_i}(0))$, it will make the canonical form
transform into
\begin{eqnarray}
(E,J_{n_i}\left( \frac{\lambda_i}{1 + z_2 a_{i0}} \right),A''_{i})
\; ,
\end{eqnarray}
where $A''_i =  f^{(z_2)}_{[(EA)A]}\circ f^{(z_2)-1}_{[(EA)J]} \circ
J_{n_i} \left( \frac{\lambda_i}{ 1+ z_2 a_{i0}} \right) $. The
entanglement classes parameterized by $\vec{\mathbf{a}} =
(\lambda_i, a_{in})$ are not inequivalent with different parametric
values, while the parameters undergo a transformation
$\mathcal{P}_{z_2}$ characterized by $z_2$, i.e.,
\begin{eqnarray}
\left.\begin{array}{l} \lambda_i \\  \\ a_{in} \end{array}\right\}
\xrightarrow{\displaystyle \mathcal{P}_{z_2}} \left\{
\begin{array}{l} \displaystyle \frac{\lambda_i}{1 + z_2 a_{i0}}
\\ \\ a''_{in}
\end{array} \right.\; .
\end{eqnarray}
Here, $a''_{in}$ are elements of $A_{i}'' $.

\subsection{The third superposition $T_{[JA]}^{(z_3)}$}
\label{sec-TJA}

In this case
\begin{eqnarray}
J_{n_i} + z_3 A_{i} =
g^{(z_3)}_{\scriptscriptstyle[JA]}(J_{n_i}(0)).
\end{eqnarray}
From the above equation it can be inferred that there should exist a
similarity transformation
\begin{eqnarray}
S^{-1} g^{(z_3)}_{\scriptscriptstyle[JA]}[J_{n_i}(0)]S =
J_{n_i}(\lambda + z_3 a_{i0}) \; .
\end{eqnarray}
Then
\begin{eqnarray}
S^{-1}J_{n_i}(0)S =
g_{\scriptscriptstyle[JA]}^{(z_3)-1}[J_{n_i}(\lambda + z_3a_{i0})]
\end{eqnarray}
and
\begin{eqnarray}
S^{-1}A_{i}S  & = & S^{-1}\cdot \, f_{[A_i]}(J_{n_i}(0)) \cdot \, S
\nonumber \\ & = & f_{[A_i]} \circ
g_{\scriptscriptstyle[JA]}^{(z_3)-1} \circ J_{n_i}\left( \lambda +
z_3 a_{i0}\right)\; .
\end{eqnarray}

We note that when the $T$ transformation of $ J_{n_i} + z_3 A_i$
applies on
\begin{eqnarray}
(E,J_{n_i}(\lambda_i),A_{i}) \; ,
\end{eqnarray}
where $A_i = f_{[A_i]}(J_{n_i}(0))$, it will make the canonical form
transform into
\begin{eqnarray}
(E,J_{n_i}\left( \lambda_i + z_3 a_{i0} \right),A'''_{i}) \; ,
\end{eqnarray}
where $A'''_i = f_{[A_i]} \circ g_{\scriptscriptstyle[JA]}^{(z_3)-1}
\circ J_{n_i}\left( \lambda + z_3 a_{i0}\right) $. The entanglement
classes parameterized by $\vec{\mathbf{a}} = (\lambda_i, a_{in})$
are not completely inequivalent with different parametric values,
while parameters undergo a transformation $\mathcal{P}_{z_3}$
characterized by $z_3$, i.e.,
\begin{eqnarray}
\left.\begin{array}{l} \lambda_i \\  \\ a_{in} \end{array}\right\}
\xrightarrow{\displaystyle \mathcal{P}_{z_3}} \left\{
\begin{array}{l} \displaystyle \lambda_i + z_3 a_{i0}
\\ \\ a'''_{in}
\end{array} \right.\; .
\end{eqnarray}
Here, $a'''_{in}$ are elements of $A_{i}'''$.

\subsection{Results}

As far we know that different $A$ are inequivalent up to the
following transformations
\begin{eqnarray}
A_{i}' & = & f^{(z_1)}_{[(EJ)A]}\circ f^{(z_1)-1}_{[(EJ)J]} \circ
J_{n_i}\left(\frac{\lambda_i}{ 1+ z_1\lambda_i}\right) \; ,
\\ A_{i}' & = &  f^{(z_2)}_{[(EA)A]} \circ f^{(z_2)-1}_{[(EA)J]} \circ
J_{n_i} \left(\frac{\lambda_i}{1 + z_2 a_{i0}}\right)\; ,
\\ A_{i}' & = & f_{[A_i]} \circ
g_{\scriptscriptstyle[JA]}^{(z_3)-1} \circ J_{n_i}\left( \lambda +
z_3 a_{i0}\right) \; .
\end{eqnarray}
or their combinations in proper orders.
\begin{eqnarray}
\begin{pmatrix}
E \\ J' \\ A'
\end{pmatrix} = \mathcal{P} \begin{pmatrix}
E \\ J \\ A
\end{pmatrix} \; ,
\end{eqnarray}
where $\mathcal{P} = \prod_{z} \mathcal{P}_{z}$, $A = \oplus A_i$.

Notice that the above-explained classification scheme should be
applicable to the $2\times N\times N$ case, a special case of the
general $L\times N\times N$ system. By then there will be only the
superposition $T^{(z_1)}_{[EJ]}$, and $(E,J)$ is just the
entanglement class up to the symmetry \cite{2nn}
\begin{eqnarray}
\lambda_i \; \xrightarrow{\displaystyle \mathcal{P}_{z_1}} \;
\displaystyle \frac{\lambda_i}{1 + z_1\lambda_i} \; ,
\end{eqnarray}
which is a special case of Eq.(\ref{first-superposition}).

\subsection{Examples}

In the following we employ the new method to characterize the
entanglement classes of $2\times N\times N$ and $3\times N\times N$
state under SLOCC. First, we show how the present method works by
comparing it with the previous result \cite{2mn} for a $2\times
N\times N$ system (Appendix A in \cite{2mn}); then we use the
present method to construct the canonical form for the specific
$3\times N\times N$ system discussed above.

\subsubsection{$2\times N\times N$ case}

The quantum state of $2\times N\times N$ can be represented by a
matrix pair $\Psi = \left(\Gamma_1, \Gamma_2 \right) $. The
invertible operations (SLOCC operations) act on the three partite in
the following forms:
\begin{eqnarray}
\Psi' = T\otimes P \otimes Q \; \Psi = T \begin{pmatrix} P \Gamma_1 Q\\
P\Gamma_2 Q \end{pmatrix} \; , \label{example-222}
\end{eqnarray}
where $T$ is a $2\times 2$ matrix. Suppose one typical canonical
form of an entanglement class is
\begin{eqnarray}
\Psi = \left( E , J \right) = \left(
\begin{pmatrix}
\ddots &   &   &   & \\   & 1 & 0 & 0 &   \\   & 0 & 1 & 0 &   \\
 & 0 & 0 & 1 &   \\ &   &   &   & \ddots
\end{pmatrix} ,
\begin{pmatrix}
\ddots &   &   &   & \\   & \lambda_i & 1 & 0 &   \\   & 0 & \lambda_i & 1 &   \\
 & 0 & 0 & \lambda_i &   \\ &   &   &   & \ddots
\end{pmatrix} \right) \; .
\end{eqnarray}
Because of the requirement $r(E) > r(J)$, the admissible $T$ matrix
that keeps the rank inequality should be in the form of an upper
triangular
$\begin{pmatrix}
t_{11} & t_{12} \\
0 & t_{22}
\end{pmatrix}$ \cite{2nn}. For the sake of convenience we take the pair $(\Gamma_1,\Gamma_2)$ to
represent the two $3\times 3$ submatrices hereafter. After the $T$
transformation
\begin{eqnarray}
\Gamma'_1 =
\begin{pmatrix}
t_{11} + t_{12}\lambda_i & t_{12} & 0  \\
0 & t_{11} + t_{12}\lambda_i & t_{12}   \\
0 & 0 & t_{11} + t_{12} \lambda_i
\end{pmatrix} ,\
\Gamma'_2 =
\begin{pmatrix}
t_{22}\lambda_i & t_{22} & 0   \\
0 & t_{22} \lambda_i & t_{22}   \\
0 & 0 & t_{22}\lambda_i
\end{pmatrix}\; ,
\end{eqnarray}
the canonical form corresponding to it is now (i.e.,
$P=Q^{-1}\Gamma_1'^{-1}$)
\begin{eqnarray}
\Psi' & = & \left( P\Gamma'_1Q , P\Gamma'_2Q \right) = (E,J') \nonumber \\
& = & \left(
\begin{pmatrix}
\ddots &   &   &   & \\   & 1 & 0 & 0 &   \\   & 0 & 1 & 0 &   \\
 & 0 & 0 & 1 &   \\ &   &   &   & \ddots
\end{pmatrix} ,\
\begin{pmatrix}
\ddots &   &   &   & \\
& \frac{t_{22} \lambda _i}{t_{11}+t_{12} \lambda _i} & 1 & 0 &   \\
& 0 & \frac{t_{22} \lambda _i}{t_{11}+t_{12} \lambda _i} & 1 &   \\
 & 0 & 0 & \frac{t_{22} \lambda _i}{t_{11}+t_{12} \lambda _i} &   \\
 &   &   &   & \ddots
\end{pmatrix} \right) \; . \label{example-2nn-prev}
\end{eqnarray}
One can regard this as a symmetry among parameters $\lambda_i \sim
\lambda'_i = \frac{t_{22} \lambda _i}{t_{11}+t_{12} \lambda _i}$
induced by the transformation $T$ in canonical forms. In the
Appendix A of \cite{2mn}, this symmetry is expressed as
\begin{eqnarray}
(E,J) \xrightarrow{T,P,Q} (E,J') \; , \label{result-2mn}
\end{eqnarray}
where the parameters in $(E,J')$ are $\lambda'_i = \frac{ t_{22}
\lambda_i }{ t_{11} + t_{12}\lambda_i}$ and thus the canonical forms
with different values of the parameter actually belong to the same
SLOCC class, that is, $\lambda_i \sim \lambda'_i$.

By employing the method developed in this work, the quantum state
can be manipulated as follows:
\begin{eqnarray}
\Psi = \left( \Gamma_1 , \Gamma_2 \right) = \left( 1 , \lambda_i +
x\right) \; , \label{cano-2nn}
\end{eqnarray}
where $x$ represents $J_{n_i}(0)$. The transformation $T$ makes
\begin{eqnarray}
\left( \Gamma'_1 , \Gamma'_2 \right) = \left( t_{11} +
t_{12}\lambda_i + t_{12}x \;,\;  t_{22} \lambda_i + t_{22}x
\right)\; .
\end{eqnarray}
The canonical form now can be obtained just by dividing the
polynomial $t_{11} + t_{12}\lambda_i + t_{12}x$ [the same role as
$\Gamma_{1}'^{-1}$ before Eq.(\ref{example-2nn-prev})]:
\begin{eqnarray}
\left( P\Gamma'_1Q, P\Gamma'_2Q \right) = \left( 1\; ,\; \frac{
t_{22} \lambda_i + t_{22}x }{t_{11} + t_{12}\lambda_i + t_{12}x}
\right)\; . \label{example11}
\end{eqnarray}
Here $ P\Gamma'_2Q$ can be expanded as
\begin{eqnarray}
\frac{ t_{22} \lambda_i + t_{22}x }{t_{11} + t_{12}\lambda_i +
t_{12}x} = \frac{t_{22} \lambda _i}{t_{11}+t_{12} \lambda
_i}+\frac{t_{11} t_{22} }{\left(t_{11}+t_{12} \lambda
_i\right){}^2}x - \frac{t_{11} t_{12} t_{22} }{\left(t_{11}+t_{12}
\lambda _i\right)^3}x^2\; .
\end{eqnarray}
Note that here $J_{n_i}(0)^3=0$. The the Jordan form of the second
matrix in Eq.(\ref{example11}) is $\frac{t_{22} \lambda
_i}{t_{11}+t_{12} \lambda _i} + J_{n_i}(0)$, and the canonical form
is then
\begin{eqnarray}
\Psi' = \left(1\; ,\; \lambda'_i + x \right) = \left(1\; ,\;
\frac{t_{22} \lambda _i}{t_{11}+t_{12} \lambda _i} + x \right) \; .
\end{eqnarray}
Compared to Eq.(\ref{cano-2nn}) we see that different parameters
$\lambda_i$ and $ \lambda'_i = \frac{t_{22} \lambda
_i}{t_{11}+t_{12} \lambda _i}$ belong to the same entanglement class
induced by the transformation $T$, which is in agreement with the
result of Eq.(\ref{result-2mn}).

\subsubsection{$3\times N\times N$ case}

We consider the case of $\psi = (A_1, A_2, A_3)$ where $r(A_{1})=N$.
Here, the $\psi$ can always be transformed into $\psi'=(E, A'_2,
A'_3)$ by invertible operators $P,Q$. The actual problem discussed
in Sec. \ref{sec-3NN} is to construct the canonical form of $(E,
A'_2, A'_3)$ under $T, P, Q$, where $[A'_2, A'_3]=0$. With our
method, this can be decomposed into two tasks: (1) the simultaneous
similarity transformation of matrix pairs $(A'_2, A'_3)$ (canonical
form) and (2) application of superpositions induced by $T$ among
$(E, A'_2, A'_3)$ (symmetry between the canonical form).

In the triple-matrix form $(E, A_2, A_3)$, its canonical form under
$P,Q$ is $(E,J,A)$. The effective $T$ transforms $(E, J, A)$ into
$(E + z_{1}J + z_{2}A, J + z_{3} A, A)$ whose canonical form is
$(E,J',A')$. That is, $T$ induces a symmetry (equivalent relation)
between different canonical forms $(E,J,A)$ and $(E,J',A')$. Our
task is to find the relation between parameters in $J'$, $A'$ and
$J$, $A$. The initial canonical form can be represented by a
polynomial form, i.e.,
\begin{eqnarray}
& & \psi = (E, J, A ) \nonumber \\ & = & \left(
\begin{pmatrix}
\ddots &   &   &   & \\   & 1 & 0 & 0 &   \\   & 0 & 1 & 0 &   \\
 & 0 & 0 & 1 &   \\ &   &   &   & \ddots
\end{pmatrix} ,
\begin{pmatrix}
\ddots &   &   &   & \\   & \lambda_i & 1 & 0 &   \\   & 0 & \lambda_i & 1 &   \\
 & 0 & 0 & \lambda_i &   \\ &   &   &   & \ddots
\end{pmatrix}, \begin{pmatrix}
\ddots &   &   &   & \\   & a_{i0} & a_{i1} & a_{i2} &   \\   & 0 & a_{i0} & a_{i1} &   \\
 & 0 & 0 & a_{i0} &   \\ &   &   &   & \ddots
\end{pmatrix} \right) \nonumber \\ & = &
\left(1 \; , \; \lambda_i + x \; , \; a_{i0} + a_{i1}x + a_{i2}x^2
\right) \; ,
\end{eqnarray}
where $x$ represents $J_{n_i}(0)$.

In the following we show that the transformation $T^{(z_1)}_{[EJ]}$
in Sec. \ref{TzEJ} may transform the above quantum state into the
following form:
\begin{eqnarray}
\left(1 + z_1\lambda_i + z_1 x \; , \; \lambda_i + x \; , \; a_{i0}
+ a_{i1}x + a_{i2}x^2 \right)\; .
\end{eqnarray}
We transform the first matrix into a unit matrix by dividing the
polynomial $1 + z_1\lambda_i + z_1 x$ and obtain the polynomial
functions $f^{(z_1)}_{[(EJ)J]}(x)$ and $f^{(z_1)}_{[(EJ)A]}(x)$.
That is,
\begin{eqnarray}
& & \left(1 , \frac{\lambda_i + x}{1 + z_1\lambda_i + z_1 x},\frac{
a_{i0} + a_{i1}x + a_{i2}x^2}{1 + z_1\lambda_i + z_1 x} \right)
\nonumber \\ & = & \left(1, \frac{\lambda_i }{1+z_1 \lambda_i
}+\frac{x}{(1+z_1 \lambda_i )^2}-\frac{z_1 x^2}{(1+z_1 \lambda_i
)^3},\right. \nonumber \\ & & \left.\frac{a_{i0}}{1+z_1 \lambda_i }
+ \frac{(a_{i1}-a_{i0} z_1+ a_{i1} z_1 \lambda_i ) x}{(1+z_1
\lambda_i )^2} +\right.\nonumber \\ & & \left.
\frac{a_{i2}(1+z_1\lambda_i)^2 - z_1(a_{i1} - a_{i0}z_1 +
a_{i1}z_1\lambda_i)}{(1+z_1 \lambda_i )^3}x^2\right) \nonumber \\ &
= & \left(1\; , \; f^{(z_1)}_{[(EJ)J]}(x) \; , \;
f^{(z_1)}_{[(EJ)A]}(x)\right) \; ,\label{example-func}
\end{eqnarray}
where only up to $x^2$ is needed due to $J^3_{n_i}(0)=0$. And, then
we transform the second matrix into Jordan form, which means the
similarity transformation
\begin{eqnarray}
M \; f^{(z_1)}_{[(EJ)J]}(x)\;M^{-1} = J_{n_i}(\frac{\lambda_i
}{1+z_1 \lambda_i } ) \; .
\end{eqnarray}
In polynomial form it reads
\begin{eqnarray}
\frac{\lambda_i }{1+z_1 \lambda_i } + \frac{y}{(1+z_1 \lambda_i
)^2}-\frac{z_1 y^2}{(1+z_1 \lambda_i )^3}  = J_{n_i}(\frac{\lambda_i
}{1+z_1 \lambda_i } ) = \frac{\lambda_i }{1+z_1 \lambda_i } + x\; ,
\label{example-333}
\end{eqnarray}
with $y=MxM^{-1}$, where $x=0$ implies $y=0$. Hence, the quantum
state is now in the form
\begin{eqnarray}
\left(1, \;J_{n_i}(\frac{\lambda_i }{1+z_1 \lambda_i } )  , \;
f^{(z_1)}_{[(EJ)A]}(y)\right)
\end{eqnarray}
correspondingly. One can solve $y$ from Eq.(\ref{example-333}) by
expressing $y = \sum_{i=1}^{\infty}b_ix^{i}$, like
\begin{eqnarray}
f^{(z_1)-1}_{[(EJ)J]}(J_{n_i}(\frac{\lambda_i }{1+z_1 \lambda_i } ))
= y = \left(1+z_1 \lambda _i\right)^2 x + z_1\left(1+z_1 \lambda
_i\right){}^3 x^2 + \cdots \; .
\end{eqnarray}
Take $y$ into $f^{(z_1)}_{[(EJ)A]}(y)$ and keep those terms up to
$x^2$, for the third matrix $A_i'$ we have
\begin{eqnarray}
A'_i & = & a'_{i0} + a'_{i1}x + a'_{i2}x^2 \nonumber \\ & = &
\frac{a_{i0}}{1 + z_1 \lambda _i} + \left(a_{i1} - a_{i0} z_{1}+
a_{i1} z_{1} \lambda _i\right) x + a_{i2} \left(1 +z_{1} \lambda
_i\right){}^3 x^2\; .
\end{eqnarray}
Finally, by comparing coefficients of different powers of $x$, we
have
\begin{eqnarray}
& & \left( \lambda_i , a_{i0} , a_{i1} , a_{i2} \right) \sim \left(
\lambda'_i , a'_{i0} , a'_{i1} , a'_{i2} \right) \nonumber \\ & = &
\left( \frac{\lambda_i}{1+z_1\lambda_i} , \frac{a_{i0}}{1 + z_1
\lambda _i}, \left(a_{i1} - a_{i0} z_{1}+ a_{i1} z_{1} \lambda
_i\right) , a_{i2} \left(1 +z_{1} \lambda _i\right){}^3 \right) \; .
\label{example-tejz1}
\end{eqnarray}
Similarly, we can get the result for $T^{(z_2)}_{[EA]}$, i.e.,
\begin{eqnarray}
& & \left( \lambda_i \; , \; a_{i0} \; , \; a_{i1} \; , \; a_{i2}
\right) \sim \left( \lambda'_i \; , \; a'_{i0} \; , \; a'_{i1} \; ,
\; a'_{i2} \right) \nonumber \\ & = & \left(
\frac{\lambda_i}{1+z_2a_{i0}} \; , \; \frac{a_{i0}}{1 + z_2 a_{i0}}
\; , \; \frac{a_{i1}}{1+a_{i0}z_2 - a_{i1}z_2\lambda_i} \; , \;
\frac{a_{i2} \left(1+a_{i0} z_2\right){}^3}{\left(1+z_2
\left(a_{i0}-a_{i1} \lambda _i\right)\right)^3} \right) \;
,\label{example-teaz2}
\end{eqnarray}
and $T_{[JA]}^{(z_3)}$
\begin{eqnarray}
& & \left( \lambda_i \; , \; a_{i0} \; , \; a_{i1} \; , \; a_{i2}
\right) \sim \left( \lambda'_i \; , \; a'_{i0} \; , \; a'_{i1} \; ,
\; a'_{i2} \right) \nonumber \\ & = & \left( \lambda_i +z_3 a_{i0}
\; , \; a_{i0} \; , \; \frac{a_{i1}}{1+a_{i1}z_3} \; , \;
\frac{a_{i2}}{\left(1+z_3 a_{i1}\right)^3} \right) \; .
\label{example-tjaz3}
\end{eqnarray}
More specifically, take Eq.(\ref{example-tjaz3}) as an example. If a
quantum state in the canonical form has one block with parameters
$\left( \lambda_i \; , \; a_{i0} \; , \; a_{i1} \; , \; a_{i2}
\right) = (1,0,2,3)$, then it is SLOCC equivalent to the canonical
form which has the corresponding block of $\left( \lambda'_i \; , \;
a'_{i0} \; , \; a'_{i1} \; , \; a'_{i2}
\right)=(1,0,\frac{1}{1+2\cdot z_3}, \frac{3}{(1+2\cdot z_3)^3})$.
In order to complete the classification, in some special cases,
i.e., when rescaling the whole matrix of $J$ or $A$ as discussed
before Eq.(\ref{T=PLDU}), zero appears in denominators in the
fractions of Eqs.(\ref{example-teaz2}) and (\ref{example-tjaz3}),
and the states have to be treated separately. In all these cases,
the method of similarity transformation plus the polynomial form of
symmetries shall be applied iteratively.

\section{Conclusion}

In this work, we propose a general classification scheme for
entangled quantum system $L\times N\times N$. In this scenario,
based on the commutativity of the ILOs $T$, $P$, $Q$, the
classification procedure is decomposed into two steps: the
simultaneous similarity transformation of a commuting matrix pair
into a canonical form and application of the internal symmetry of
parameters in the canonical form. This is innovative to the general
entanglement classifications, because by this scheme we can always
extract symmetry properties from the general equivalent relation and
then leave a relative simple canonical form for the entanglement
class. In most cases the major challenge comes from the detailed
forms of the representation of symmetries. For demonstration, a
concrete example of entanglement classification for a type of
$3\times N\times N$ system is presented, of which the symmetry
properties are expressed in forms of polynomial functions.

\vspace{0.7cm} {\bf Acknowledgments}

This work was supported in part by the National Natural Science
Foundation of China(NSFC) and by the CAS Key Projects KJCX2-yw-N29
and H92A0200S2. We thank Bin Liu for initial collaboration on this
work.

\newpage

\appendix{\bf\Large Appendix}

\section{The structure of matrix $A$} \label{structure-matrix-A}

\noindent {\bf I}. If the Jordan form $J = \oplus
J_{n_i}(\lambda_i)$ is nonderogatory, that is, every $\lambda_i$ has
geometric multiplicity 1, then matrices commuting with it can be
expressed as $A = \oplus_i A_i$, where $A_{i}$ is in the form of
Eq.(\ref{A_i-form}) (see, for example, theorem {\bf S2.2} in
\cite{Matrix-Polynomials}).

\noindent {\bf II}. If some of the Jordan blocks of $J$ have the
same eigenvalue,
\begin{eqnarray}
J = \cdots \oplus\underbrace{J_{n_1}(\lambda_1)\oplus
J_{n_2}(\lambda_2)\oplus \cdots \oplus
J_{n_l}(\lambda_l)}_{\lambda_1=\lambda_2=\cdots=\lambda_l}\oplus
\cdots \; ,\label{derogatory}
\end{eqnarray}
where $n_i\geq n_j$ for $i>j$. (We can always gather the Jordan
blocks with same $\lambda_i$ together.) The matrix commuting with
this block can be expressed as
\begin{eqnarray}
A' = \begin{pmatrix} A_{11} & A_{12} & \cdots & A_{1l} \\ A_{21} &
A_{22} & \cdots & A_{2l} \\ \vdots & \vdots & \ddots & \vdots \\
A_{l1} & A_{l2} & \cdots & A_{ll}
\end{pmatrix} \; , \nonumber
\end{eqnarray}
where $A_{ij}$ is an $n_i\times n_j$ upper triangular Toeplitz
matrix. As the main subject of our work is the study of parametric
symmetries, we refer to Eqs. (S2.13$-$S2.15) in
\cite{Matrix-Polynomials} for the detailed definition of $A_{ij}$.
Here we give an example of $n_1 = 3, n_2=2$:
\begin{eqnarray}
\begin{pmatrix} A_{11} & A_{12} \\
A_{21} & A_{22}
\end{pmatrix} = \begin{pmatrix}
a_{110} & a_{111} & a_{112} & a_{120} & a_{121} \\ 0 & a_{110} &
a_{111} & 0 & a_{120} \\ 0 & 0 & a_{110} & 0 & 0 \\ 0 & a_{210} &
a_{211} & a_{220} & a_{221} \\ 0 & 0 & a_{210} & 0 & a_{220}
\end{pmatrix}\; . \label{matrix-A}
\end{eqnarray}
This can be represented  phenomenologically by the following matrix
of polynomials:
\begin{eqnarray}
\begin{pmatrix}
f_{[A_{11}]}(x) & f_{[A_{12}]}(x) \\ f_{[A_{21}]}(x) &
f_{[A_{22}]}(x)
\end{pmatrix}
= \begin{pmatrix} a_{110} + a_{111}x + a_{112}x^2  & a_{120} +
a_{121}x \\ a_{210}x + a_{211}x^2  & a_{220} + a_{221}x
\end{pmatrix}\; , \label{polynomial-A}
\end{eqnarray}
where $x = J_{n_{1}}(0)$ given that $n_1\geq n_2$, and the last
$n_{1}-n_2$ rows and columns of Eq.(\ref{polynomial-A}) should be
omitted to give the form of Eq.(\ref{matrix-A}). In the discussion
of the parametric symmetry of the canonical form in Secs.
\ref{sec-TEA} and \ref{sec-TJA}, the supposition and inverse
functions applied in Eq.(\ref{polynomial-A}) can be proceeded
directly with the only difference being the emergence of the
nondiagonal elements, i.e., $f_{[A_{12}]}$ and $f_{[A_{21}]}$ in
Eq.(\ref{polynomial-A}).

\end{document}